\pdfoutput=1
\documentclass[11pt, a4paper, oneside]{Thesis} 

\graphicspath{{Pictures/}} 

\usepackage[square, numbers, comma, sort&compress]{natbib} 

\usepackage{amsmath}
\usepackage{amsfonts}
\usepackage{amssymb}
\usepackage{graphicx}
\usepackage{physics}

\usepackage[T1]{fontenc} 
\usepackage{natbib}
\usepackage{setspace}
\title{\ttitle} 

\begin{document}

\frontmatter 

\setstretch{1.3} 

\fancyhead{} 
\rhead{\thepage} 
\lhead{} 

\pagestyle{fancy} 

%

\thesistitle{Real World Evaluation of Approaches to Research Paper Recommendation}
\documenttype{Undergraduate Thesis}
\supervisor{Dr. Joeran Beel}
\supervisorposition{Asst. Prof.}
\supervisorinstitute{Trinity College Dublin, Ireland}
\cosupervisor{Mr. Tirtharaj Dash}
\cosupervisorposition{Lecturer}
\cosupervisorinstitute{Birla Institute of Technology and Science}
\examiner{}
\degree{Master of Science (Hons.) Economics \& Bachelor of Engineering (Hons.) Computer Science}
\coursecode{BITS F422}
\coursename{Thesis}
\authors{Siddharth Sankaran Dinesh}
\IDNumber{2012B3A7519G}
\addresses{}
\subject{}
\university{\texorpdfstring{\href{http://universe.bits-pilani.ac.in/} 
                {Birla Institute of Technology and Science}} 
                {Birla Institute of Technology and Science}}
\UNIVERSITY{\texorpdfstring{\href{http://universe.bits-pilani.ac.in/} 
                {BIRLA INSTITUTE OF TECHNOLOGY AND SCIENCE, PILANI}} 
                {BIRLA INSTITUTE OF TECHNOLOGY AND SCIENCE, PILANI}}
\CAMPUS{\texorpdfstring{\href{http://bits-goa.ac.in} 
                {K. K. BIRLA GOA CAMPUS}} 
                {K. K. BIRLA GOA CAMPUS}}

\department{\texorpdfstring{\href{http://www.bits-pilani.ac.in/goa/ComputerScienceInformationsSystems/ComputerScienceandInformationSystems} 
                {Computer Science}} 
                {Computer Science}}
\DEPARTMENT{\texorpdfstring{\href{http://universe.bits-pilani.ac.in/hyderabad/mechanicalengineering/MechanicalEngineering} 
                {MECHANICAL ENGINEERING}} 
                {MECHANICAL ENGINEERING}}
\group{\texorpdfstring{\href{Research Group Web Site URL Here (include http://)}
                {Research Group Name}} 
                {Research Group Name}}
\GROUP{\texorpdfstring{\href{Research Group Web Site URL Here (include http://)}
                {RESEARCH GROUP NAME (IN BLOCK CAPITALS)}}
                {RESEARCH GROUP NAME (IN BLOCK CAPITALS)}}
\faculty{\texorpdfstring{\href{Faculty Web Site URL Here (include http://)}
                {Faculty Name}}
                {Faculty Name}}
\FACULTY{\texorpdfstring{\href{Faculty Web Site URL Here (include http://)}
                {FACULTY NAME (IN BLOCK CAPITALS)}}
                {FACULTY NAME (IN BLOCK CAPITALS)}}

\maketitle
\Certificate

\begin{abstract}
In this work, we have identified the need for choosing baseline approaches for research-paper recommendation systems. Following a literature survey of all research paper recommendation approaches described over the last four years, we framed criteria that makes for a well-rounded set of baselines. These are implemented on Mr. DLib a literature recommendation platform. User click data was collected as part of an ongoing experiment in collaboration with our partner Gesis. We reported the results from our evaluation for the experiments. We will be able to draw clearer conclusions as time passes. We find that a term based similarity search performs better than keyword based approaches. These results are a good starting point in finding performance improvements for related document searches.
\end{abstract}

\begin{acknowledgements}

I am grateful for the good health and well being that were necessary to complete this thesis.

I wish to express my sincere thanks to Prof. Joeran Beel for hosting me at the ADAPT Center at Trinity College Dublin, Dublin, and providing a high level of guidence for my undergraduate thesis as well as for providing me with all the necessary facilities for the research.

I place on record, my sincere thanks to my co-guides Mr. Tirtharaj Dash and Dr. Mridula Goel, for all the guidance and support in my undergraduate research. 

I am also grateful to my colleagues at Mr. DLib: Sophie, Stefan and Sara. I am extremely thankful to them for sharing expertise, and sincere and valuable guidance and encouragement extended to me.

I also thank my parents for the unceasing encouragement, support and attention.

I am also thankful for the funding provided through a Science Foundation of Ireland Grant to ADAPT center which supported my stay in Ireland.
\end{acknowledgements}


\pagestyle{fancy}

\lhead{\emph{Contents}} 
\tableofcontents 

\lhead{\emph{List of Figures}}
\listoffigures 

\lhead{\emph{List of Tables}}
\listoftables 







\clearpage
\thispagestyle{empty}

{\Huge \bfseries Notations \& Conventions}


\begin{itemize}
    \item CTR - Click Through Rate
    \item Recommendation systems, recommender systems, recommendation approaches are all interchangeably used in a loose sense. By any of these, I mean to refer to an abstract function that takes as input a user's preferences and returns as output a set of recommended items that are supposedly relevant to the user.
    \item Requested document is sometimes used. By this I mean the document for which we are providing recommendations. 
    \item CF - Collaborative Filtering
    \item CBF - Content Based Filtering

\end{itemize}











\mainmatter 

\pagestyle{fancy} 



\chapter{Background} 

\label{Chapter1} 

\lhead{Background} 


\section{Introduction}

Research is a continuous process. The purpose of research is to introduce new ideas through scientific discourse. As more and more journal articles and conference papers are published year by year, it becomes increasingly difficult to identify research articles that are related to one’s field of interest. Furthermore, it becomes non-trivial to keep up-to-date with newly published research articles as well as to associate them to previously published articles. 

With the digitization of research publications, there has been a move to use computers to augment the search for related articles which are relevant to a researcher’s field of interest. Such systems are known as research paper recommendation systems. A recommender system can be taken as a black box which takes in a profile of a user and matches it against a candidate set of items in order to suggest previously unseen items for a user. These items are considered to be the most relevant recommendations for that user.

\section{Recommender Systems}

As explained previously, a recommender system can be most easily visualized as a system that takes as input some characteristics from a user which are processed in order to identify items which are most relevant to the user’s interests. The type of matching used commonly categorizes the approach into either a content based approach, or a collaborative filtering approach. 

In a content based filtering approach, the tastes and interests of a user are extracted by using the information contained in the items that the user has previously interacted with. The exact action that is considered as an interaction depends on the specifics of the recommender system. For example, a book recommender system might choose to use the act of purchasing a book as an interaction, whereas, a friend recommender system on a social media platform might choose to use the act of sending a message as a relevant interaction. The items that a user interacts with are usually summarized in an \textit{item profile} and this item profile is then compared against the candidate set of items to provide personally tailored recommendations. 

In a collaborative filtering approach, no information from the items that a user interacts with is used. Instead, the similarities are considered between the users in terms of the items that each user has interacted with. For a new recommendation for a user, a set of most similar users is considered. Then items that similar users have interacted with, but the user for whom we are providing recommendations has not interacted with are used as recommendations. 

In order to improve the performance of the recommenders, there have been successful efforts to combine the traditional collaborative filtering approach with a traditional content based filtering approach. These type of recommenders are said to use Hybrid recommendation approaches. One way to create a hybrid approach is to use separately construct recommendation lists and then combine them. Another way is to add collaborative filtering ideas into a content based filtering framework, and vice-versa. 

As examples, Pandora Radio\footnote{www.pandora.com} is a content based music recommender system. IMDb\footnote{www.imdb.com} is a content based movie recommender system. Spotify\footnote{www.spotify.com} has a collaborative filtering music recommender system. Netflix\footnote{www.netflix.com} is a good example of a movie and TV show recommender system that combines the features of traditional content based filtering and traditional collaborative filtering approaches. 

\section{Research Paper Recommendation}

Research-paper recommendation addresses the task of providing recommendations based on an abstraction of the user's profile. More than 200 research articles regarding research-paper recommendation systems have been published in the 16 years until 2015, and there have been more new systems introduced since then which will be described in chapter \ref{Chapter 2}.

Depending on the type of information that is available to the research paper recommendation system, we can ascertain if the system uses a collaborative filtering approach or a content based filtering approach. Usually, if a user in the recommendation system has a library of literature articles associated to him, there is a possibility that the system uses a collaborative approach. However, if this is not the case, and there is no library associated to a user, we can comfortably conclude that the system uses a content based approach. Additionally, a system may or may not use the history of literature that a user interacted with in order to recommend new articles. 

Literature recommendation is usually found to be associated with a reference management software, or a digital library. Usually collaborative filtering approaches would only provide new recommendations when it is used with intervals of time between recommendations. Thus, there is a distinction between a \textit{related article} feature in digital libraries which only uses a content based approach and a \textit{weekly recommendation} feature in a digital library which would probably use a collaborative approach. 

In this work, the focus is only on identifying baselines for research paper recommendations for a related article search feature. Thus, we will look primarily at content based recommenders. 

A typical content based research paper recommendation framework is illustrated in Fig. 1. Research paper recommendation is a specific domain in the larger field of recommender systems. The items are research articles, and users are researchers who need to identify relevant literature that pertains to their specific area of interest.

\begin{figure}
    \centering
    \includegraphics[scale=0.9]{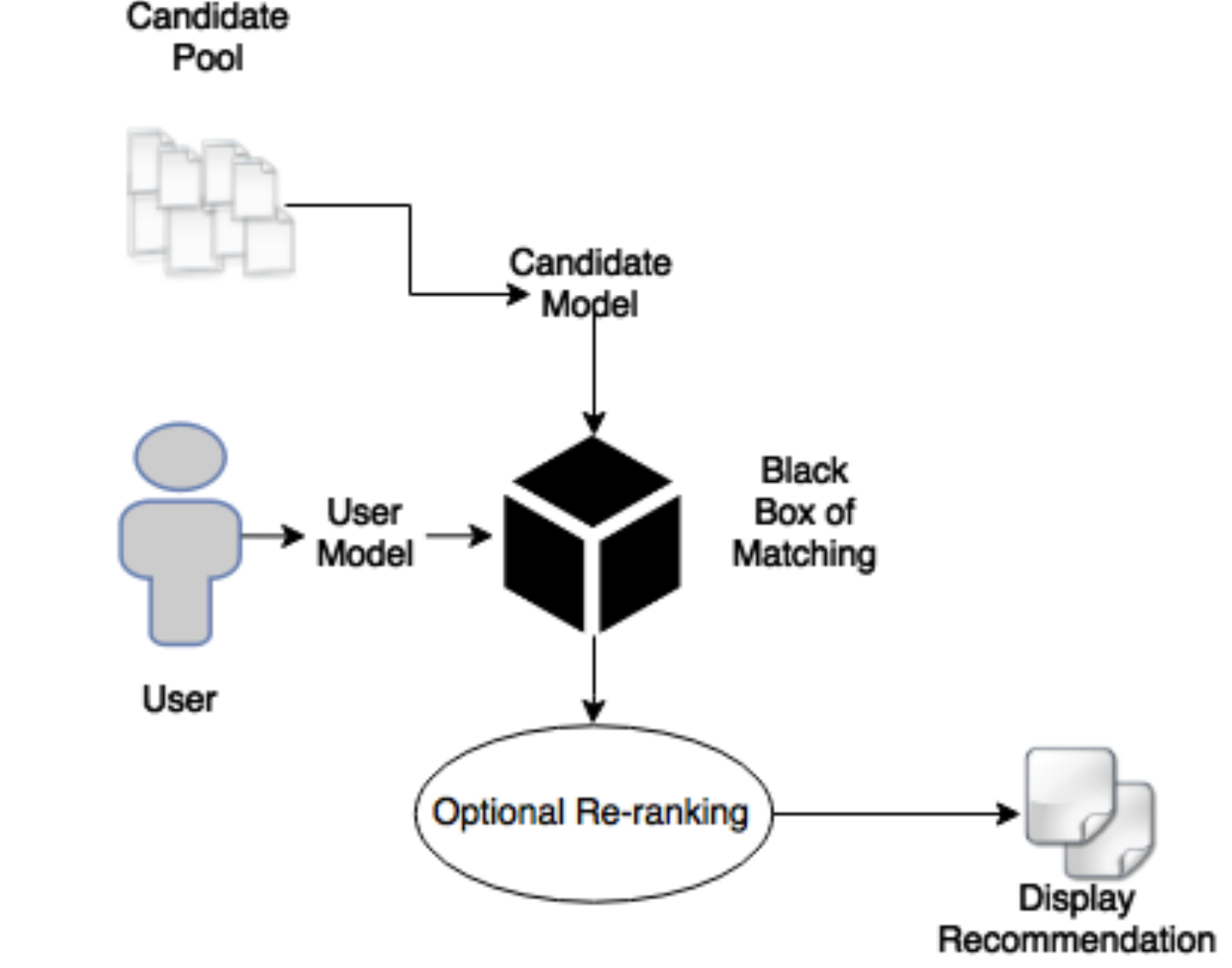}
    \caption{Schematic of a Research Paper Recommender Framework}
    \label{fig:my_label}
\end{figure}

The framework creates candidate models for research articles in the candidate pool.


\chapter{Literature Review} 

\label{Chapter 2} 

\lhead{Chapter 2. \emph{Lit. Review}} 

In 2015, Beel et. al. published an extensive literature survey of the field of research paper recommender systems until and including the year 2013\cite{Beel2015rlitreview}. In order to pick the papers that they included in the survey, they used the following query:
"[paper | article | citation] [recommender | recommendation] [system | systems]" 
The query was used as a search input to Google Scholar, ACM Digital Library, Springer Link, and ScienceDirect systems. In the literature review for this thesis, we used this same query.
Beel et. al. identified 217 articles in total and described the trends in the field based on the categories based on which evaluation methods were applied (e.g. user-studies or offline evaluations), which evaluation metrics were used (e.g.precision or recall), and the scale of the evaluation as well as the dataset or document catalog used for evaluations.

The following sections provide a summary of the work published in the field of research paper recommender systems since the end of 2013 up till August 2016. It is organized according to the a very broad categorization of what kind of recommendation approach is used.

\section{Collaborative Filtering}

\subsection{Personalized Academic Paper Recommendation System}
Lee et. al. \cite{lee2015personalized} used the collaborative filtering approach to construct individual recommendations for researchers. They use a simple k-nearest neighbors approach to find similar users to the target user and make recommendations for the user. They consider all the papers that a user has authored in order to construct their user model. The user is modelled by a text vectorization of his papers and simple vector cosine is used as the similarity measure. 

The evaluation is carried out in an online and offline fashion. The offline evaluation dealt with measuring the classification accuracy of the recommendations on a corpus of IEEE Xplore and ACM Digital Library papers. The offline method primarily evaluates whether the paper recommendations made fall under the same field of research that the user is focused in.  There is no focus on serendipity of recommendations. The online test was a user study with three participants, the results of which cannot be representative of the method. Regardless, their methodology shows a high user satisfaction among the participants. 

\section{Content Based Filtering}

\subsection{Sugiyama and Kan}

Sugiyama and Kan  published updates to their previous work in research paper recommendation systems over three articles since 2013 \cite{Sugiyama:2015:THR:2719943.2719947} \cite{Sugiyama:2013:EPC:2467696.2467701} \cite{Sugiyama2015}. All their work is explained in \cite{Sugiyama2015}. Their approach has been to recommend papers based on an author’s profile, where the author is the user in their system. In order to get serendipitous results, they linearly combine the user model of the author with the user models of one or many dissimilar users. They also brought in the idea of discovering potential citation papers in addition to the already cited papers in the article in order to expand the contextual network for a paper under consideration. Citation context around every citation in the paper is also a feature that they use. Each of these papers is modelled by its terms in order to get the user profile, and compared against the word model of the candidate papers.

They evaluate their model on the ACM Digital library with the ground truth being constructed based on input from 50 researchers. The dataset has been made available publicly. Their model is shown to be superior to some state of the art models such as Wang and Blei’s topic modelling approach and Nascimento’s approach in terms of recommendation accuracy. They also compare the serendipity of their models against Carbonell and Goldstein’s maximal marginal relevance model  as well as against a random model to show that the nITN measure (which is not referenced anywhere else in their text) is significantly better than the competing methodologies. 

\subsection{Effective Academic Research Papers Recommendation for Non-profiled Users}

Hanyurwimfura proposes a solution to recommend scientific articles to non-profiled users in their 2015 work \cite{hanyurwimfura2015effective}. This methodology is meant to avoid the problems of collaborative filtering for users for whom there is not enough data available to build their user profile. They take a content based approach in extracting both short and long queries from a single paper provided as the input. The long queries are taken from the abstract and sections similar to the title whereas the short queries are commonly occurring phrases in the paper as well as words from the title. These queries are weighted and used to filter candidate papers from the corpus. The recommendations are made using a simple cosine similarity between the target paper and the filtered papers.

They evaluate their approach in the sense of a topic extraction paradigm and also as a recommendation system. The evaluation of the relevancy of the topics extracted is done by 20 participants who are researchers and the average acceptance ratio was found to be 68.3\%. To evaluate the performance as a paper recommendation tool, one paper from each researcher was used to build recommendations and each recommendation had to be rated for its relatedness to their field of work. They reported recall and NDCG scores using a corpus of ACM, IEEE and Science Direct as being incrementally better than using the methods of Nascimento et. al. \cite{nascimento2011source} 

\subsection{SimSeerX}

SimSeerX is a similar paper search engine built on top of the CiteSeer document corpus by Kyle Williams as part of Lee Giles’ research group \cite{Williams2014}. It can take whole documents or text as a query and returns a list of the most similar documents to the query. The idea behind SimSeerX’s structure is to decompose a document into a set of signatures. Each document is then indexed into the system by these signatures. A document submitted as a query is also decomposed into its signature and then searched against the rest of the indexed document. Any similarity function can be used in this system because the document signatures have been built to accommodate for this. Currently, SimSeerX supports Key Phrase Similarity, Shingle Similarity, and something known as a SimHash similarity by customizing the similarity functions in the Solr instance which has indexed all the documents.

Since the paper was an exposition of the system, the evaluation was done in terms of the time taken for a query to return results, both in a cold-start and a cached scenario. It was shown that the system scales well upto 3.5 million document corpus size. 

\subsection{PaperTaste}

The authors, Xue et. al., aim to solve recommendation as a supervised ranking problem \cite{xue2014personalized}. They split the corpus into two parts based on a time-frame. The older papers form the training set and the new ones are the validation/test set. A sample in the training data would be an input paper and its citation network combined with randomly sampled uncited papers (which constitute the negative samples). The scores for each node is the number of times that paper has been cited, which is 0 for the negative samples. The problem then becomes the issue of finding suitable features to train this ranking model. 

The authors choose to construct features such as the page rank for paper, author and venue, the age of the paper, content similarity between titles, abstracts etc. They also use a feature which they call the “author ratio”. It is the percentage of papers in the user’s profile which contain some or all authors of the candidate paper. Using these features, they train a Ranking SVM model. 

A recommendation for a new paper is made by constructing these features for each candidate paper. If serendipity is a requirement, the authors rank the top 500 recommendations and randomly sample from that list. 

Evaluation was done against a few baseline approaches such as a basic CF, basic CBF and a Page Rank weighted CF method. In the offline evalution, which was done on a Social Scholar dataset of 730,605 papers for 10,000 authors, it was reported that PaperTaste system outperformed the others in terms of the NDCG$_{k}$ value. Their online evaluation only constituted the reporting of statistics about the percentage of people who activated the recommendation option in their PaperTaste system and then proceeded to interact with the system. 

\subsection{Recommendation System Based on Fuzzy Cognitive Map}

Their method augments the key words present in an input paper to find a fuzzy match between these key words and words in a pre-prepared ontology of topics in the research domain \cite{liu2014recommendation}. Their paper is a bit tough to read in terms of grammar, but they go on to describe something known as a Fuzzy Cognitive Map between topics in the ontology to build recommendations. Their evaluation is done using user studies in comparison with the RecULike system. Not much information is given on how this evaluation was conducted. 

\subsection{Keyword Based Article Recommendation System using Map-Reduce}

In their 2015 paper, Singh and Ahuja \cite{singh2015article} provided a proof of concept in the utilization of Hadoop based technologies to provide paper recommendations. It was not complicated to implement, and their mechanism only does simple keyword matching from the input query. They provide an evaluation in terms of the time taken per number of queries to show that using Hadoop based map reduce infrastructures is essential for large scale recommendations. 

\subsection{Content-Based Approach in Research Paper Recommendation System}

Philip and others in a 2014 paper \cite{philip2014application} use a keyword based vector space model to make article recommendations for digital libraries. They build a system with user interactions in order to build a user profile. They model papers by their keywords using a tfidf approach and uses the cosine similarity measure to find relevant articles to recommend based on an input query. No evaluation of their framework was provided in this paper.

\subsection{RefSeer}

In 2014, Wenyi Huang authored an article \cite{Huang:2014:RCR:2740769.2740832} that encapsulates all the work done so far in building the RefSeer system as part of C. Lee Giles’ research group. The paper describes the system behind RefSeer as well as giving a concise description of the recommendation methodology used. RefSeer recommends citations for queries. Usually the queries are uncited manuscripts of paragraphs of text. First, to create the global recommendations, Refseer finds the topics in the corpus using the Cite-PLSA-LDA model \cite{huang2012recommending}. For a new query, the top 5 topics are calculated and recommendations are made for these topics. Only with the addition of a local recommendation, does the recommendation system become personalized. 

This is done by a local recommendation which makes use of a translation model between the original manuscript and the possible citation papers. This model is learned from the corpus using an IBM-1 model with pairs of phrases from the original text and the citations for those texts. For a fresh query, the system makes use of these two frameworks to come up with the list of citations needed.

The Refseer team evaluates their work on the CiteSeer corpus and the CiteULike corpus and report the training and recommendation times. The training time is quite sizeable on the entirety of the CiteSeer dataset. However the recommendation time for a fresh query is less than 5 milliseconds. They also report that their MRR on the datasets, but there is no comparison with regard to any other system provided, with an acknowledgment that their model is not superior to other approaches to paper recommendation. 

\subsection{Recommender Systems with Big Data}

This paper \cite{jokar2016contextual} published in 2016 presents a recommendation system that uses cosine similarity between a user profile and keywords, abstracts of articles to suggest recommendations. The user profile is non-dynamic and is a function of the user’s working area. The system was evaluated through a user study with recommendations being suggested based on a corpus of IEEE documents. It does not add anything to the field. 

\subsection{Investigating the User Curriculum}

Magalhaes et. al. built a recommendation system in 2015 that harnesses the vast repository of research articles in Portuguese language database CV-Lattes \cite{magalhaes2015recommending}. The model papers based on terms and concepts. The papers are then indexed based on these. Each paper is associated to concepts by different weights. The user profile is modelled by features taken from the CV-Lattes system. Their experiments deal with finding out how the length of the user profile (in years) affects the performance of the recommendations. Their experiments also focused on the comparability of the paper’s being modelled by terms as opposed to concepts present in the papers. They also compared against Lopes’s recommendation methodology \cite{Lopes:2008:PRS:1666091.1666103}. They report that their system outperformed the existing system, however, using more information than Lopes’ system. They do not compare the system against more state of the art baselines. 

\subsection{Recommendations using User's Preferences}

Igbe et. al. adapt a frequent pattern growth algorithm in order to prune out a set of recommended papers from a larger set of candidate research articles \cite{igbe2016incorporating}. To do this, they first build and extract features from each paper’s metadata. These features all have values between 0 and 1. Many of these features are based on the paper’s citation data. The authors then compare the average feature score against the cut-off feature score in order to limit the scope of the recommendations. Only he articles which have average feature scores above the cutoff are taken into consideration for the next step. For the next step, the authors take user input to select an optional number of search filters to prune the intermediate set of research papers. The frequent pattern growth algorithm is used in this step to select as many papers as possible which satisfy as many filters as needed to satisfy a pre-specified minimum support value. These are the final recommendations.

The user inputs are keywords and an optional number of search filters. The authors use an offline evaluation to study their approach in comparison to two other baselines. However the baselines are the Page Rank model, and Sugiyama and Kan’s 2010 model \cite{sugiyama2010scholarly}, both of which have further been refined to produce improvements. This means that the improvements as reflected in the paper may not be an improvement on the current state of the art versions of these models. Additionally the paper does not describe the corpus that the authors have used for their comparisons. 

\subsection{Personalized Concept-Driven Recommender System for Scientific Libraries}

De Nart et. al. prototyped the idea of assigning keywords to all scientific articles in the corpus using the Dikepe Keyphrase extraction module \cite{Nart2014APC}. These keywords are represented in the form of a context graph to cluster similar keywords together. Linked keywords form a network of contexts. They represent the user profile in terms of keywords of the papers that the user rates as relevant. This approach does not overcome the cold-start problem and is not much different from many existing systems. The prototype was rated using user studies by 30 graduate students after using it for a period of a month. They also evaluate the approach on the MovieLens dataset for movies. 

\subsection{Science Concierge}

This system was developed as a recommendation system to recommend research articles which are presented in one particular conference \cite{achakulvisut2016science}. The recommendations are not meant to be serendipitous, instead, the aim is to recommend research articles which are as close to the desired topic. The distance is judged using a human curated topic hierarchy. 

The idea behind Science concierge is the vectorization of keywords and abstracts of research papers using Latent Semantic Analysis. The system then calculates the nearest neighbors of a set of input research papers. Since nearest neighbor search is expensive, the search is approximated using ball trees. In terms of performance, and using the evaluation criteria mentioned earlier, the Achakulvist et. al. report the scores in the paper as generally better than when only keywords are considered to represent the document.

\subsection{PubRec}

Alzoghbi’s content based model represents both the research articles and user profiles in terms of domain related keywords \cite{DBLP:conf/lwa/AlzoghbiA0L15}. Varying weights are given to keywords extracted from different sections of the text in terms of the relative importance of the sections. The researcher profile is uniquely built by use of a multivariate regression in terms of his previous publications. The weights on the past publications are decayed based on the age of the article. Then a simple dot product of the article vector and the research profile vector gives the similarity between the article and the user’s tastes, according to the authors. 

The experiment conducted by Alzaghbi et. al. was to recommend interesting papers for 50 researchers. These results were compared versus Nascimento’s \cite{nascimento2011source} and Sugiyama’s \cite{sugiyama2010scholarly} earlier works in terms of MRR and NDCG. Their reported results reveal that PubRec comfortably outperforms Nascimento’s work (based on keywords) and is competitive to Sugiyama’s recent works in the field. The experiments were conducted using Sugiyama’s Scholarly publication recommendation dataset. 

\subsection{Neural Probabilistic Model for Context Based Citation Recommendation}

Huang et. al. as part of Lee Giles’ group developed a neural probabilistic model to learn a semantic embedding to represent research papers \cite{Huang:2015:NPM:2886521.2886655}. The model that they build eventually learns the probability of citing a paper given the citation context. The query to the model is a citation context, and the output is a list of possible documents to cite. This is done by projecting both the documents and the citation contexts into a shared embedded vector space. Prior to this, word representations and document representations are learned separately. In the final step, batches of pairs of citation contexts and their citation documents are used as training sample to train the neural probabilistic model. 

In the experiment they conducted, they use the CiteSeer dataset and split it into a training period and a testing period. Their aim was to predict the citations and compare it against other models such as the Citation Translation Model (CTM) \cite{huang2012recommending} and the Cite-PLSA-LDA model \cite{huang2012recommending}. They compared the models in terms of the MRR, MAP and NDCG values. Their results conclusively showed the usefulness of this method. It outperformed every model except for the CTM by 2 or 3 folds. CTM was outperformed by a few percentage points as well. It is not clear whether this model has been incorporated into the CiteSeer system. 

\subsection{LDA-Based Approach}

Amami et. al. used an LDA approach to build the author profile \cite{DBLP:conf/nldb/AmamiPSF16}. Assuming that the authors publications are representative of his interests, they construct the author profile by applying LDA to the abstracts of the author’s publications. Each candidate paper for recommendation is then brought down to a language model (i.e., a set of topics it contains). An important step in the construction of the user profile is the validation of the topic model using a hold-out set of published articles.

The similarity between the author profile and the candidate paper is calculated in terms of KL divergence. KL divergence is a similarity measure that is used between two probability distributions. In this case, both the author profile and the candidate paper are represented by a probability distribution over the topics present in the texts. 

This approach is evaluated against Wang and Blei’s  Collaborative Topic Regression model\cite{wang2011collaborative} and Zhang’s CAT model\cite{Zhang2014}. Recommendations are made on the ArnetMiner dataset\footnote{https://cn.aminer.org/} which has around 1.5 million papers. However, this set of papers is whittled down and recommendations are only considered for 1600 authors. The ground truth for each of the papers is the reference list for the paper. The results reported are only for the recall at varying recommendation lengths. The model outperforms the compared models for all number of recommendations in terms of its recall capacity.  

\section{Graph Based Filtering}

\subsection{Authoritative scholarly paper recommendation based on paper communities}

Zhou et. al. built an authoritative paper recommendation system which was to help junior researchers identify the important papers in their field of study \cite{zhou2014authoritative}. This was done by identifying communities in the citation network of the entire corpus using the Greedy Clique Expansion algorithm. Within each community, the Paper Rank algorithm was used to rank the most influential nodes in the community. These nodes are then suggested as the most authoritative papers in that community. 

Their evaluation seemed like a verification of their procedure in that they compared the rankings within five sample communities in the HEP-TP corpus against the number of citations from within that community. No comparison against other ranking algorithm was done, and there was no information as such to combine the rankings between communities to provide, say, a top-K list of recommendations similar to the query paper. 

There is mention of using the input papers from the researcher in a diagram, so I suppose that they will identify one or two communities relevant to the author and run Paper Rank in those communities only, thus reducing the search space. 

\subsection{Common Author Authority Propagation (CAAP)}

Hsiao et.al. propose a methodology \cite{hsiao2015model} to recommend highly authoritative papers which are related to a query document by making use of the common author network of that paper. Their CAAP method uses authority propagation in the citation network of that query paper as a backup in case the Common Author Network fails to yield any valid recommendations. Their Common Author approach is a Google Scholar search for publications by the co-authors. Out of these, using the terms extracted from the title and keywords, candidate papers are compared for similarity to try to find authors who have published a chain of 2-3 papers related to the same topic. If any such chain is found, then the Common Author approach is successful and is provided as a recommendation, otherwise the Author Propagation in the citation network is used as a backup. This is done on a citation network constructed out of querying the Google Scholar engine for related articles to the terms extracted out of the query paper. Authority propagation serves to filter out the unimportant nodes in the citation network. 

Evaluation was done in an offline fashion where Recall and Maximum Average Precision were calculated. The results were only compared against a Unified Graph Model (UGM) \cite{Meng:2013:UGM:2541167.2507831}. The experiments showed that the CAAP approach outperformed the UGM on 300 research papers from a few of the top conferences in the Computer Science field.

\subsection{Query-oriented Approach for Relevance in Citation Networks}

Totti’s work \cite{totti2016query} deals with giving recommendations for an input query. His approach involves an initial content based search to get a number of similar documents and then expanding this candidate set by including into the candidate set all the cited and citation papers that are ‘H’ hops away in both directions. A citation network is built within this subset of papers with each edge being weighted by a combination of text similarity, query similarity and an age decay factor. However, these edge weights are only used for the IQRA-MC model which uses a random walk approach to identify the most influential nodes in this network. Upon testing, Totti reports that a simpler approach of just recommending the top cited articles within this network yields better recommendations. He calls this the IQRA-TC approach. 

The evaluations were conducted on a subset of the CiteSeerX dataset with 657,000 publications. Offline evaluations were conducted and MAP as well as NDCG were reported in comparison to a number of baseline approaches including Page Rank, Google Scholar, and Arnet Miner. The test set was created by domain experts by choosing 20-30 papers from the dataset for each query which was to be evaluated. Experiments were separately conducted for a subset of this dataset to evaluate the performance on survey papers and using the paper title as the query itself. They report that the IQRA-TC model outperforms all the other approaches on this gold-standard dataset. The authors have also made this dataset available publicly. 

\subsection{DiSCern}

Chakraborty et. al. developed a citation recommendation system \cite{chakraborty2015discern} in 2013, where they take a search query as the input. They find relevant articles related to the search query using the keywords listed in those papers and construct a citation graph between these articles. Then using the DiSCern algorithm, they find the most influential nodes in this graph. DiSCern algorithm is a variation of a vertex-reinforced random walk approach where the transition probabilities between nodes is dynamic over the course of the iterations. However, Chakraborty et. al. report that there is no appreciable improvement in precision or recall between their algorithm and common alternatives such as Page Rank. They do however say that DiSCern gives a much more diverse recommendation set as compared to other approaches, which might help with serendipitous recommendations. 

Their experiments were conducted on the Microsoft Academic Research Dataset and a High Energy Physics Dataset. For these offline experiments they compare DiSCern generated rankings with PageRank using both Relevancy metrics and Diversity metrics. DiSCern only shows improvements in the Diversity metrics such as l-hop graph density and l-expansion ratio, which are both described in the paper.

\subsection{Ferosa}

Chakraborty et. al. then created a paper recommendation system after their work with DiSCern in 2016 \cite{chakraborty2016ferosa}. Ferosa not only recommends papers but it categorizes the papers by tags based on which parts of the recommended papers are the basis for the recommendation for the given input paper. These tags are: alternate approach, background, methods, experiments etc. This is useful because it allows the researcher to quickly search for the kind of recommendations they want. These type of recommendations are labelled as Faceted recommendations by Chakraborty in his paper. 

Based on the initial input paper, a network is created by considering all the citation and cited papers. Then using a random walk approach with restarts, a subset of this network is combined with papers that have content similarity with the target article to generate the final recommendations. This is done for each tag separately. However, in order to compare against existing recommendation systems, they also use an aggregation method to select the overall top recommendations in a method called r-Ferosa. 

Their evaluation is done on the AAN corpus of papers using user studies. They measure the Overall Precision and Overall Impression. They compare against Google Scholar, Microsoft Academic Search and another graph based approach proposed by Liang et. al \cite{DBLP:conf/waim/LiangLQ11}. The authors present that the user studies show results favoring the r-Ferosa framework of paper recommendations. An evaluation was also conducted to show the effectiveness of the faceted recommendations in comparison with two other self-constructed baselines as the authors claim that no other faceted recommendation system currently exists to compare against.

\subsection{ClusCite}

ClusCite is a citation recommendation system that operates on the entire corpus at once \cite{ren2014cluscite}. The approach taken is to cluster the citation network of the entire corpus into soft clusters and then to learn a recommendation function for each of these clusters separately. This is done by extracting meta-information from the graph as features and optimizing an equation which derives the paper relative authority within each cluster group. When a query paper is submitted, they classify this new paper into one of these clusters and using the devised function, identify the relevant papers to be recommended. 

Experiments are conducted on the Pub-Med\footnote{\url{https://github.com/shanzhenren2/PubMed\_subset}} and DBLP corpuses\footnote{\url{http://arnetminer.org/DBLP\_Citation}} and they show substantial improvement to existing approaches in an offline evaluation. In their evaluation, Ren et. al. report performance improvements against systems like rank-SVM  \cite{Nie:2005:ORB:1060745.1060828}, LDA approach, Link-PLSA-LDA \cite{Nallapati:2008:JLT:1401890.1401957}, and authority propagation \cite{Joachims:2002:OSE:775047.775067}. The authors also present results from a case study to emphasize how their approach is superior to the other compared approaches by listing the most authoritative venues, authors and papers in a few sample clusters. Their approach, being a citation recommendation system, is lacking in the area of serendipitous recommendations.

\subsection{Exploiting Social Relations}

Huynh et. al. developed a method to augment the users’ professional relations in the academic field to improve research paper recommendations \cite{huynh2016exploiting}. They reduced the task of recommendations into the task of training a function from the domain of the product of the set of papers and researchers into the range of a ranked list of papers. The method they have developed is offline, which means that the training time is all that is considered. Once the recommendations are made, they are stored in memory and made available when needed to the users. Their approach focuses on extracting features from the citation and co-author network in the academic graph. 

To study their approach, they made use of the Microsoft Academic Search corpus and computed the recommendations for a 1000 authors. They mention that the training time is large, which is why they picked only 1000 authors to demonstrate their procedure. They compared against their results against the results from Sugiyama’s work \cite{Sugiyama:2015:THR:2719943.2719947} in an offline evaluation. For this they prepared the dataset by selecting the papers published prior to 2006 as the training set and the papers published after that as the test set. Their results show that this “Trend Trust” approach is slightly superior to Sugiyama’s work, and vastly superior to the vanilla baseline approaches of Content Based and Collaborative filtering.

\subsection{BABEL: EigenFactor Recommends}

BABEL is a web application developed by Wesley-Smith et. al., released in 2016 and aims to be a platform for researchers to test out new algorithms for paper recommendations \cite{wesley2015experimental}. It was developed using the SSRN social science corpus, but has tools to use other corpuses as well. The platform captures metrics such as the click through rate in an attempt to quantify the performance of the experimental algorithms which are hosted on the platform. One of these experimental algorithms is EigenFactor Recommends, developed and tested by the authors. This method involves harnessing the citation graph to calculate Eigenvector centrality in each sub-topic of the corpus at an article level. This way, key papers within each sub-field are identified in an offline fashion. Their method has two variants. One optimizes for authoritative papers, and the other optimizes for serendipitous recommendations. 

The authors evaluated the system for a week by tracking the system’s click through rate (CTR) for both the eigenfactor variants against a control of the co-download recommendation system. Their results show that the co-download rate performed 4-5 times better than either of the two variants in terms of CTR.

\subsection{Academic Paper Recommendation Based on Heterogeneous Graph}

Pan et. al. convert the recommendation problem into a graph similarity problem in this work \cite{pan2015academic}. They use the AAN corpus for their research. They construct a citation graph and a word-word similarity graph (using WordNet). The two graphs are interconnected using the tfidf scores for the words in the papers. The authors then use graph similarity to calculate a similarity score between every two papers in the dataset, which is a computationally expensive task. Their final recommendations are the n-most similar recommendations to the target paper.

Because the process is computationally expensive, the authors only evaluate this method using 15 input papers in an offline evaluation. They compare their results against other graph based models such as CC-IDF, Co-citation, HITS and a purely content based method. Their approach performs substantially better in terms of NDCG and MRR scores, however as explained before, the process is computationally infeasible for large scale recommendation. 

\section{Hybrid Recommendation Systems}

\subsection{AHITS-UPT}

Devised by Lu Meilian et. al., AHITS-UPT stands for Advanced Hyperlink Induced Topic Search with User Paper Topic network \cite{meilian2015ahits}. It was developed with the aim of giving serendipitous recommendations with content based filtering. AHITS is an improvement to an iterative algorithm (HITS) which is meant to assign author and paper values to each author and paper in a network given the constraints of the network. This is more or less a graph ranking algorithm (one that is meant to identify important nodes in the graph). 

Their framework is a hybrid model because after using LDA to find topics of the papers an author is interested in, they find similar users who have similar topic profiles. For this, they have kept track of user data. If among these users, there are topics which are not explored by the target user, the system finds the top papers from the top authors in those topics. Finally, the top recommendations are the most similar (in terms of cosine similarity) between the target paper and the candidate papers. 

The system was evaluated in three different ways. A time complexity analysis was done. There was a comparison done between the authoritative author rate and the high quality paper rate. Finally the authors compared the accuracy of the AHITS-UPT recommendation system in an offline fashion. The expereiments were done on a Microsoft Academic Search dataset comprising of 160,000 articles for 10 unique users over 829 user interactions. The comparison was against the HITS recommendation method, the MHITS recommendation method and a traditional content based filtering method. Although in the final comparison, the content based filtering method was left out in the results. They show that the AHITS method was superior to the other HITS based methods, as well as being computationally less expensive. 

\subsection{Document Recommender Agent Based on Hybrid Approach}

This approach presented by Chekima et. al. is a simple hybrid model in the sense that it divides the recommendations between a collaborative approach and a content based approach which is based on bigrams \cite{chekima2014document}. The user profile for the collaborative filter is constructed by keeping track of all the articles that a user interacts with through the recommender agent. A recommendation is done by finding similar users who have also browsed the paper that the target user is currently browsing and considering the papers that the similar users have also read. Then the weights of these candidate papers are increased or decreased depending on the category that the paper falls into (using the ACM category keywords), by matching the bigrams from the article’s keywords with the keywords of the articles that the target user has already browsed. In case there are no similar users, Chekima’s system will recommend articles from the same ACM categories as the predominant category that most of the papers that the user has browsed falls into. His approach does not deal with the issues of a cold start, as no recommendations will be made if the user has not browsed any article yet. 

The evaluation was carried out against very basic Content based filtering and Collaborative filtering approaches to show that a hybrid model constituting a mixture of the features of the separate approaches performs better. The authors did not provide information on the dataset that they used for the above comparison. 

\subsection{Rec4LRW}

This is a complete literature review framework built by Sesagiri Raamkumar et. al. in 2015. It is meant to be an all in one tool for research \cite{raamkumar2015rec4lrw}. The framework comprises of three separate steps, which happen at three different points of the research process. Initially, when a researcher wants exposure to a new topic, he is to provide keywords and an optional list of research articles as the basis for the recommendation. This is the initial input. The framework then uses keyword similarity method to find an initial set of similar papers. This initial set of papers is re-ranked based on coverage and citation count and 20 top papers are shortlisted for the first phase. In the second phase, the original list of 20 papers is expanded by making use of the citation network and a content based recommender which utilizes the BM25 algorithm \cite{Jones:2000:PMI:364119.364120}. For the third phase, the framework takes in two additional inputs from the user: the type of research article, and the potential keywords that the user will be using in their manuscript. In combination with the final reading list of the user for that project, the third phase recommends a list of citations based on the reading list, the potential keywords and the preferred type of research article. This is done by making use of an item-based collaborative filtering approach to get the candidate set of papers. The candidate set is ranked based on textual similarity with articles in the reading list and the titles of the papers in the reading list to select the top 20 papers as citation recommendations for the project. 

The evaluation for this system was done a period of three months with user studies \cite{raamkumar2016papers}. The results were published in a follow up article in 2016. The user study consisted of 116 participants equally split between staff and students. The study reported agreement percentages for 7 qualitative attributes of the system such as Relevance, Usefulness etc. This paper is an interesting and comprehensive approach to article recommendation.

\subsection{Hybrid Parallel Approach for Personalized Literature Recommendation System}

This system prototyped by Ma et. al. addresses the issue of overcoming cold-starts in the Collaborative Filtering approach by first categorizing each document in the corpus using topic models \cite{ma2014hybrid}.  A scraper and parser collects all the documents publicly available and categorizes them into various fields. For a new user, a common stereotype set of recommendations are made. The prototype then logs user actions in order to build a user profile over time. This user profile is used in collaborative filtering to create new recommendations for the user. No evaluation of their system was provided in their work. 

\section{Baselines for Research Paper Recommendation}

In this literature review, there was no common set of approaches that were used as baselines for research paper recommender systems. So the focus was expanded to look for baselines in a more general field, and to apply the principles used in these larger fields into the sub-field.

In the broader field of Information Retrieval (IR), Muehleisen et al. illustrate how IR systems which are reported to be implementing the same baseline approach can have differing results on the same dataset \cite{Muhleisen}. The differences might be caused by different implementations of the backend, or different parameters in the algorithms that were used by the backend. Lin et al. address the gap in reproducibility by making available a repository containing all the code needed to run a set of standardized open-source IR baselines for the TREC dataset. The baselines are executed with one common script on a virtual machine to ensure reproducibility of the selected algorithms \cite{DBLP:conf/ecir/LinCTCCFIMV16}. The baselines used in their work were widely available, simple to implement, and relevant to the techniques being studied. An analogous common set of approaches is missing in the research-paper recommendation domain, to the best of our knowledge. 

Beel et al. have shown that similar recommendation approaches with only minor variations may perform vastly differently in evaluations against different datasets\cite{Beel2016z}. One way to account for the performance differences would be to use a set of baselines each of which using different recommendation ideas. For example, we could use a baseline set comprising of CBF, Collaborative Filtering, Graph based approaches. 


\chapter{Methodology} 

\label{Chapter3} 

\lhead{Methodology} 


\section{What is a Good Baseline?}

For literature recommendation, the set of baseline approaches should vary depending on the novel approach that is being evaluated. For example, it would be not be wise to evaluate a citation based approach against a stereotype and a content based approach but not against another citation based approach. For related article search, it would be ill-advised to evaluate a citation based or collaborative filtering approach which takes a user model constituting just one document. Thus, a good baseline approach should be:
\begin{enumerate}
\item \textbf{Easy to Reproduce:} The implementation of baselines to compare novel approaches against should not be a stumbling block in the process of research. The parameters in the approaches, if any, should be made clear to aid reproducibility. 
\item \textbf{Relevant to the task:} The baselines should fit the scope of the evaluation. If the novel approach is a CBF approach, it is advisable to compare it against at least one other CBF baseline, as well as other types of recommendations.
\item \textbf{System agnostic:} An approach that works for a wider set of documents, would have an advantage over an approach that has limited operability to a subset of the corpus. This means that the former approach can be used in a wider variety of recommendation systems. For example, baselines that are inherently multilingual in nature can be used both in scenarios where the document corpus is multilingual, and in scenarios where the corpus is of only one language.
\end{enumerate} 

\section{Choosing Baselines}

The baselines that we chose use only the information from the documents. We avoided the use of external information, thus allowing our suggested approaches to be compatible with other document catalogs when they are implemented on other recommendation systems.

Research paper recommendation approaches can be categorized into Content Based Filtering, Collaborative Filtering (CF), Graph Based Filtering, Stereotype  and Hybrid models. The baselines we chose have to be relevant to the task of recommending for a related article search. CF, Graph based, and Hybrid approaches were eliminated from the shortlist because they are not compatible for related article search.

\subsection{Random}

 The approach randomly picks the set of documents to recommend to the user. We experiment with this approach by randomly choosing to apply a language filter 50\% of the time. With the language filter, the recommended documents share the same language as the input document. The average time complexity of this approach is O(n), where n is the number of recommended documents.

\subsection{Lucene's MoreLikeThis (MLT)}
This is the most common \textit approach used in the comparison of compare new CBF approaches. The approach concatenates and tokenizes the title, abstract, keywords, and journal name using Apache Lucene's\footnote{http://lucene.apache.org/core} out-of-the-box \textit{Standard Tokenizer}. The tokens are then indexed, and recommendations are made using Lucene's \textit{More Like This} feature. We chose this approach because it consistently provides recommendations and can be easily implemented by researchers looking for a standard content based filtering baseline. This approach takes O($|$D$|$ * $|$V$|$), where $|$T$|$ is the size of the term vocabulary, and $|$D$|$ is the size of the document corpus.

\subsection{Stereotype}
Stereotyping uses a very primitive user modeling strategy with fixed recommendation classes. Users are classified, or stereotyped into generic groups and each group is assigned the same set of recommendations. In this study, we stereotype all users as researchers, and selected documents that are of common research interest, such as documents related to 'Academic Writing', or 'Experimental Practices'. The Stereotype approach has an average time complexity of O(1) as it is precompiled. 

\subsection{Most Popular}
The most popular research documents according to our partner's most viewed and most exported lists are provided as recommendations for the users. This approach is also a database read in our implementation and can be done in constant time.

\subsection{Key-phrase Based}
This is an advanced approach which is an adaptation of the Key-phrase based approach used by Ferrera et al.\cite{Ferrara2011} Whereas the original approach requires the full text of a paper to build acceptable key-phrases, we adapted the approach to do so even with only the title of the paper as input. This approach was parametrized by us and the best parameters were empirically found to be a similarity search using three each of unigram and trigram key-phrases computed using the title and abstract, if available.\cite{keyphrase} Using Lucene, this approach has an average case time of O($|$D$|$ * $|$K$|$), where $|$K$|$ is the size of the keyphrase vocabulary, and $|$D$|$ is the size of the document corpus.

In Table 3.1, we describe how each of our chosen baselines coheres with the criterion that we recommended in an earlier section. Each approach is rated between "Moderate, High and Very High" for the convenience of the reader.
\begin{table}
\centering
\caption{Coherence of Baselines with Criteria}
\label{my-label}
\begin{tabular}{l|l|l|l}
\textbf{Approach}                                                  & \textbf{Ease to Reproduce}                                                                                   & \textbf{Task Relevancy}                                                                                                                          & \textbf{System Agnostic}                                                                                                                                                                              \\ \hline

\textbf{Stereotype}                                                & \begin{tabular}[c]{@{}l@{}}Moderate -- \\ As long as\\ stereotype \\ classes \\ are the same\end{tabular} & \begin{tabular}[c]{@{}l@{}}High -- It's a\\ comparison against\\ a human curated \\ recommendation list\end{tabular}                    & \begin{tabular}[c]{@{}l@{}}High -- Stereotypes\\ can be created\\ for any data corpus\end{tabular} \\ \hline
\textbf{Random}                                                    & Very High                                                                                                 & Moderate                                                                                                                                & Very High                                                                                                                                                                                    \\ \hline
\textbf{Metadata}                                                  & \begin{tabular}[c]{@{}l@{}}High -- \\ Out of the box\\ Lucene setup\end{tabular}                          & \begin{tabular}[c]{@{}l@{}}High -- uses all fields\\  available for a related \\ paper search\end{tabular}                              & \begin{tabular}[c]{@{}l@{}}Very High -- Applicable\\ to all documents\\  in the corpus\end{tabular}                                                                                          \\ \hline
\begin{tabular}[c]{@{}l@{}}\textbf{Most} \\ \textbf{Popular}\end{tabular}   & \begin{tabular}[c]{@{}l@{}}High --\\ Over a period \\ of time\end{tabular}                                & \begin{tabular}[c]{@{}l@{}}Moderate -- Asks the\\ question: Are\\ recommendations\\ supposed to be\\ serendipitous or not?\end{tabular} & \begin{tabular}[c]{@{}l@{}}Very High -- Doesn't\\ depend on the\\ document for which \\ recommendations\\ are requested\end{tabular}                                                         \\ \hline
\begin{tabular}[c]{@{}l@{}}\textbf{Keyphrase}\\ \textbf{Based}\end{tabular} & \begin{tabular}[c]{@{}l@{}}High -- \\ Pipeline is \\ available on github\end{tabular}                     & \begin{tabular}[c]{@{}l@{}}High -- Related paper\\  search favors \\ CBF approaches\\  such as this one\end{tabular}                    & \begin{tabular}[c]{@{}l@{}}Moderate -- Current\\ version not \\ multilingual\end{tabular}                            
\end{tabular}
\end{table}

\section{Data Catalog}
\subsection{Description}
We apply and evaluate the suggested baseline algorithms in a recommender system, which is integrated in a digital library containing more than 9.5 million documents called Sowieport, which is hosted by Gesis. These documents have representation from at least 59 languages. Almost 40\% of the documents, around 4 million documents, have an abstract. 70\% of these abstracts, 3.3 million documents are English. 6 million documents only have a title associated with them, of which 2 million are English.\\ \\

\subsection{Fields Available}
A document in the catalog can have the following information:

Each document in the catalog contains the title, the journal it was published in, the publication year, and keywords as assigned by the author and an optional abstract.

\begin{table}
\caption{Multilingualism of the catalog}
\begin{tabular}{llr}
\hline\noalign{\smallskip}
Language & Title &	Abstract \\
\noalign{\smallskip}
\hline
\noalign{\smallskip}
English & 5,356,952 \ \ \ \ \ \ \ \ \ \ \  &	3,353,406\\
German & 2,045,562 & 641,263\\
No Language Specified & 1,470,385 & 0\\
All 57 Other Languages & 632,846
 & 3300\\
\hline\noalign{\smallskip}
Total & 9,505,745 & 3,998,029\\
\hline\noalign{\smallskip}
There are 5,667,917 documents without an abstract		\\
\hline
\end{tabular}
\end{table}

\section{Experiment}
\subsection{Baseline Comparison}
The ongoing experiments are conducted by providing recommendations based on our partner's catalogue and displaying the recommendations on our partner's digital library. We track the recommendations requests for documents and the clicks on the recommended documents. 

Online evaluations incorporate the human aspect of recommendations. Offline evalations, such as Mean Average Precision, use less information than the CTR does and are shown to correlate highly by Beel et al. \cite{Beel:2013:CAO:2532508.2532511}. We use CTR as the measure for evaluation of the baseline approaches.
\begin{equation}
    CTR =\frac{Number\ of\ Clicks\ Recorded}{Number\ of\ Recommendations\ Delivered}
\end{equation}

We compare the performance of the five baseline algorithms over a period of a month. For each request, the recommendations use one the five suggested approaches. If the input document is English, the Keyphrase approach is chosen 70\% of the time, Stereotype 4\%, Most Popular 4\%, Random 2\%, and Lucene's MLT 20\%. Whenever the keyphrase approach returns no related documents, we retry once more with Lucene's MLT approach. For non-English documents, we currently do not use the keyphrase approach and instead default to Lucene's MLT. This distribution of approaches was chosen because of the need to provide a high standard of recommendations to the users. It would not be ideal if they recieved totally random recommendations often, or the same stereotype recommendations once every 10 times. The clicks on the recommended documents are logged and used to calculate and compare the CTRs of the set of baselines. For each recommendation request, we recommend 6 documents.

\subsection{Analysis of Keyphrase Based Approach}
We conducted three different experiments using the extracted keyphrases: 
\begin{enumerate}
\item By providing recommendations using keyphrases generated from the title only as opposed to keyphrases constructed from the title and abstract of the document.
\item By using different combinations of unigram, bigram, and trigram keyphrases. For instance, we provide recommendations using unigrams and trigrams constructed from the title only. 
\item In randomly choosing the number of keyphrases we used in the similarity calculations. If we chose to use 5 keyphrases for a recommendation, we would use the 5 keyphrases with highest keyphraseness score. So, we could compare the effectiveness of recommendations provided using five unigrams and five trigrams con-structed from the title and the abstract of a document. 
\end{enumerate}

To evaluate the the recommendations that Mr. DLib provides, we use the Click-Through-Rate, which is the percentage of delivered recommendations that have been clicked by the user. All data collected will be shared on the Harvard Dataverse . All comparisons are done using a t-test and results significant at a p-value of 0.05 are reported. To create baselines to compare the performance of the keyphrase approach, we implemented two other algorithms in Mr. DLib:
\begin{enumerate}
\item We used Lucene’s MLT feature to recommend documents related to a query document. In using this feature, we indexed all textual meta-data related to a doc-ument on Lucene, such as Title, Abstract, and author-provided keywords. Related documents are then recommended using Lucene’s inbuilt BM-25 similarity metric. All other settings of the MLT function were as they were out-of-the-box in Lucene 6.3.0.
    \item 	A completely random baseline by which we selected completely random English language documents as recommendations for any query document. This was conceived as a control in the experiment.

\end{enumerate}


\chapter{Implementation} 

\label{Chapter4} 

\lhead{Implementation} 

\section{Mr. DLib - Literature Recommendation As A Service}

Mr. DLib is a machine-readable digital library\cite{beel2011introducing} which recently has been extended with a function that allows third parties to request literature recommendations\cite{beelmrjdcl} from Mr. DLib. These recommendations can then be displayed on their own websites. 

Mr.DLib currently has two partners which use the API for recommendations. The first is the Sowiport digital library, maintained by GESIS\footnote{www.gesis.org} in Germany. The recommendations are displayed on each article's page using a Javascript element which dynamically requests and loads he recommendations. This process is described further below in this section. The second partner is the reference management application called Jabref\footnote{www.jabref.org}. Mr. DLib was integrated through a related articles tab which dynamically calls the API and processes the XML returned from the query\cite{Feyer2017}. The integration was done so that a majority of control over the display of the recommendations remains with Mr. DLib. This was done by returning HTML snippets that contain the formatting as well as content about each delivered recommendation. The formatting can, of course, be changed at the end of Mr. DLib, thus allowing for live experimentation of different formatting and display options without needing to change the code in the Jabref project.  

Maintaining and building a recommender system as a service such as Mr. DLib comes with a lot of benefits and challenges\cite{DBLP:conf/ecir/BeelD17}. The benefits are, of course, the ability to simultaneously compare the performance of many different recommendation approaches in a real-world setting. The data collected can be said to be more representative than from lab experiments. Another benefit is that approaches can be tweaked based on historical performance as well as the comparitive performance of other approaches. For example, the same approach can be tested using various parameter settings to find what can be considered as the optimal setting. A third advantage is that recommendation approaches can be compared across different document corporas to see if there are changes in performance across corpora. The challenges include having to deal with extremely noisy data and problems with creating a scalable way to implement a distribution for the randomization of approaches.

Mr. DLib's architecture is depicted in Fig. 4-1.
Mr. DLib is mostly developed in JAVA and uses standard tools and libraries whenever possible.

The central element of Mr. DLib is its Master Storage, namely a MySQL database. This database contains all relevant data including documents’ metadata and statistics of delivered recommendations. Metadata of documents includes:
\begin{enumerate}
    \item  Mr. DLib’s document ID
    \item  Partner’s Document ID
    \item Title
    \item Authors
    \item Abstract
    \item Keywords
    \item Published in (generic field for journal name, conference, etc.)
    \item Language
    \item Publication  Year
    \item  Document Type (journal article, conference article, …)
\end{enumerate}

\begin{figure}
    \centering
    \includegraphics[scale=0.2]{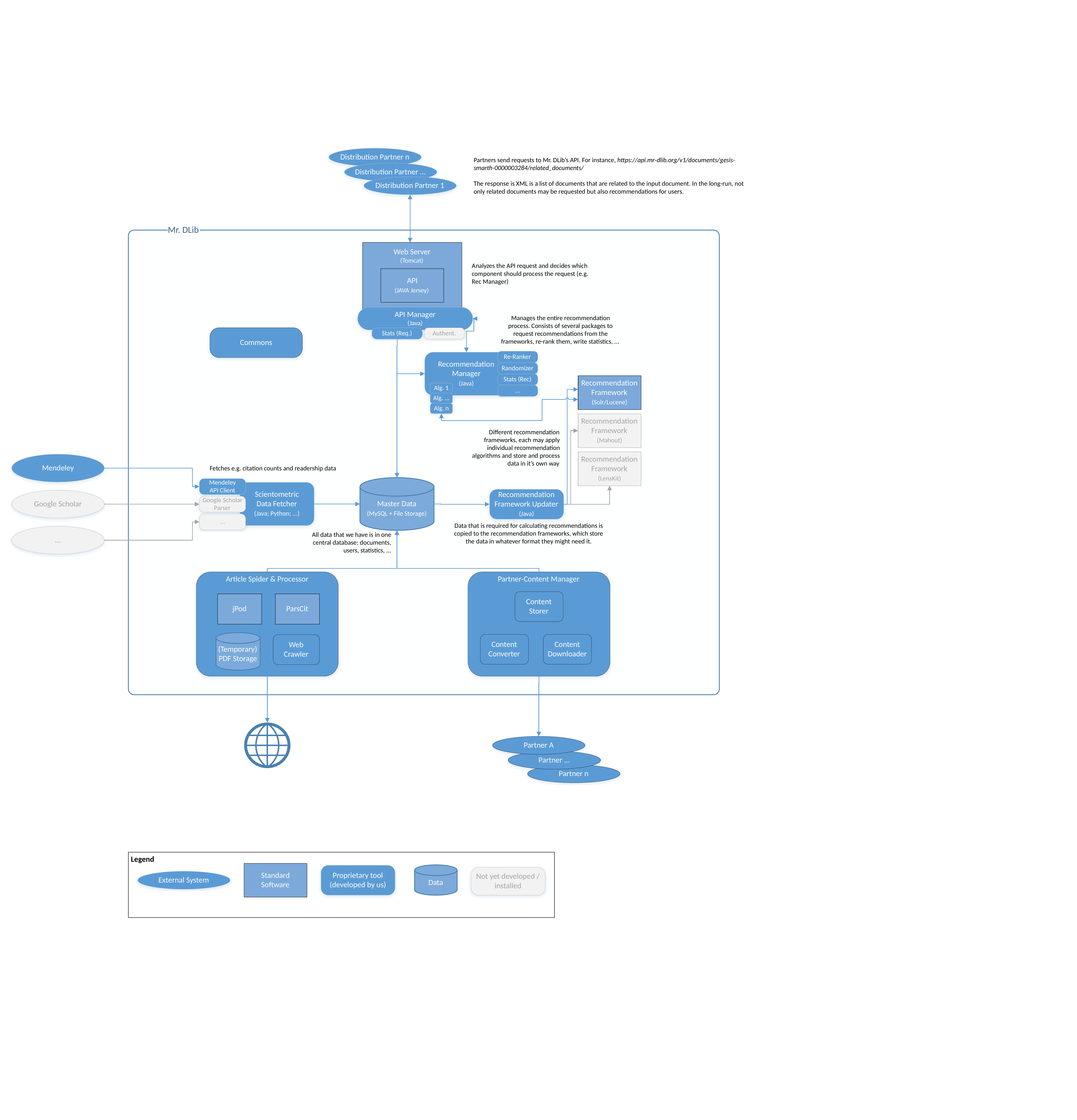}
    \caption{Mr. DLib system's architecture}
    \label{fig:4.1}
\end{figure}
The “Content Retriever” downloads the partners’ content once a month. Currently, Mr. DLib has only one partner, and this partner provides a corpus of 9.5 million documents. GESIS provides its data as Solr XML export. The XML files are backuped on Mr. DLibs server and then the relevant metadata of the documents is stored in the database. Although, GESIS provides full-texts for some documents, Mr. DLib currently does not yet utilizes it in its recommender system due to storage and CPU restraints.

To generate recommendations, Mr. DLib uses Apache Solr/Lucene as recommendation framework. Lucene offers an integrated “More like this” function that calculates content-based document similarities. This standard approach is used over the documents’ titles and abstracts to find related documents for one given input document. While currently this simple standard approach is used, it is our highest priority, to increase recommendation effectiveness. Therefore, we are currently experimenting with increasing ranking accuracy based on Mendeley Readership data; utilizing semantic modelling and applying the knowledge from our previous research. In the long-run, more than one recommendation framework will be used by Mr. DLib. Potential further recommendation frameworks are, for instance, Apache Mahout and LensKit. The prospect of using different recommendation frameworks is also the reason why we decided to have one master storage with all information. This way, every recommendation framework can retrieve the required data from this central storage. Storing the partner’s data directly from the XML files in the various recommendation frameworks would be error prone. The creators of the recommendation platform were also among the creators of the recommender system in Docear, a reference management software, so the architecture of Mr. DLib borrows from architecture of the recommender system in Docear\cite{Beel2014bdocearArch}. 

The ‘Scientometric Data Fetcher’ gathers data from external sources to enhance the recommendation process. Currently, Mr. DLib requests for each document the readership statistics from Mendeley’s API . In the future, further data such as citation counts might be fetched e.g. from Google Scholar. The readership statistics are used to re-rank recommendations based on the document’s popularity on Mendeley. Incorporating bibliometrics into a research paper recommender system is meant to identify the most reliable and important research articles from the larger set of relevant recommendations\cite{DBLP:conf/ecir/SiebertDF17}.  

Mr. DLib offers a REST API, i.e. partner may send requests as HTTP request (typically GET). To retrieve recommendations, the partner calls https://api.mr-dlib.org/v1/documents/<partner-document\_id>/related\_documents/ and retrieves an XML response containing a list of related documents (JSON is planned). Mr. DLib’s web service is realized with Apache Tomcat and JAVA Jersey. The proprietary “API Manager” writes some statistics to the database and forward the requests to the proprietary “Recommendation Manager”. 
The “Recommendation Manager” (JAVA) handles all processes related to the recommendations. It looks up required data from the database (e.g. matches the partner’s document id from the URL with Mr. DLib’s internal document ID), decides which recommendation framework to use, calculates and stores statistics, and re-ranks recommendation candidates based on scientometrics\footnote{Description of Architecture is copied from Mr. DLib's official documentation}. 

\section{Recommendations to the Digital Library}

The ongoing experiments are conducted by providing recommendations using the Sowieport catalogue and displaying the recommendations on our partner's digital library. We track the recommendations requests for documents and the clicks on the recommended documents.

As seen in Fig. 4.2, when a user accesses a document in the digital library, he is provided with all available fields pertaining to the document such as Title, Abstract, Year of Publication, Publication Journal, Keywords, and Citations. 

Additionally, on the left hand side, there are displayed a number of documents which are supposed to be related to the original document. Henceforth, we will refer to this document as the requested document. All recommended documents form a recommendation set. A recommendation set usually contains 6 related documents, and lesser if the system could not find six related documents. The number of recommendations that are to be displayed is described by the problem of choice overload and has been analyzed\cite{beierle2017exploring} as part of the Mr.DLib project. The findings from the experiment were that lower click-through rates were observed for higher numbers of recommendations and twice as many clicked recommendations when displaying ten related articles instead of one related article. Their results indicate that users might quickly feel
overloaded by choice. With this in mind, for the experiments that we conducted, we decided to use six as the size of the recommendation set.

\begin{figure}
    \centering
    \includegraphics[scale=0.5]{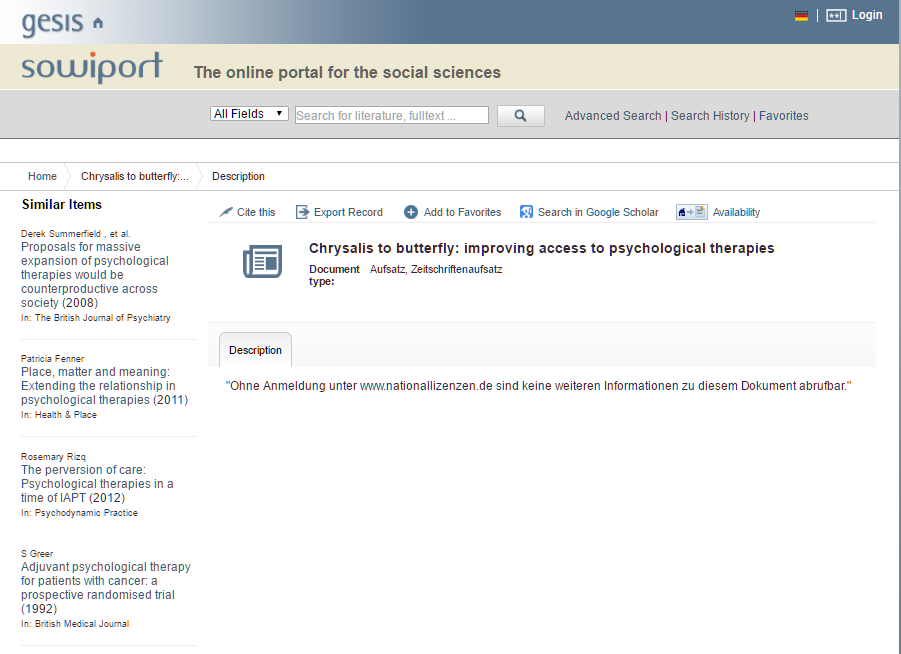}
    \caption{Landing page on Sowieport for a research article}
    \label{fig:4.2}
\end{figure}

Each time a document is accessed in the digital library, Mr. DLib's partner, GESIS, sends a recommendation request to Mr. DLib's servers. The recommendation request is processed and an XML output is returned as shown in Fig. 4.3. This XML output has all the information needed to generate the display on the side of the digital library.

\begin{figure}
    \centering
    \includegraphics[scale=0.5]{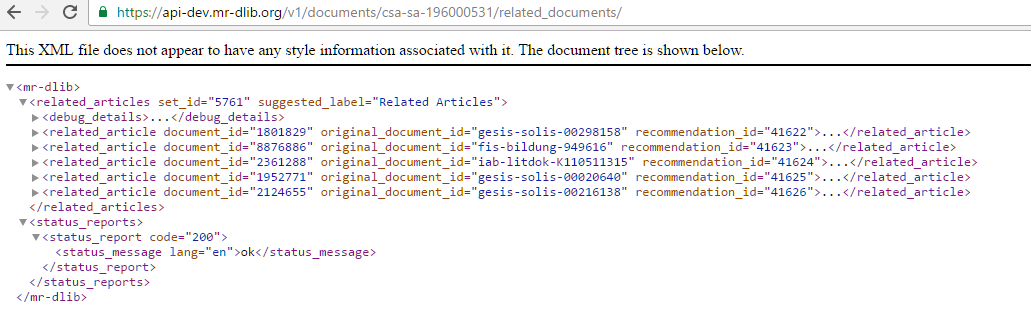}
    \caption{XML output from Mr. DLib's system}
    \label{fig:4.3}
\end{figure}

The XML output is then parsed by our partner and is rendered in a column format on the left hand side of the landing page, as seen in Fig. 4.4. When a user is interested by one of these displayed recommendations, he/she clicks on the hyperlink attached to the recommendation. This click is forwarded to Mr. DLib's servers in order to record the click in the database, and redirects to the landing page of this clicked document. The document which was clicked is then opened in a new tab, in order to allow the user the opportunity to click on other recommendations as well.
\begin{figure}
    \centering
    \includegraphics[scale=0.5]{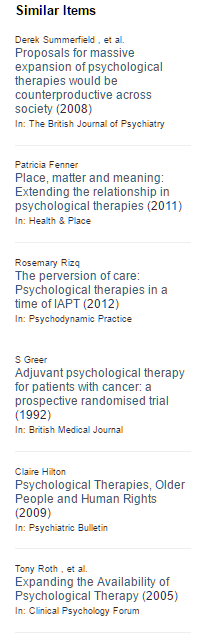}
    \caption{The XML output is parsed and displayed on the left hand side of the requested document in Sowieport}
    \label{fig:4.4}
\end{figure}

\section{Implementation of Keyphrase approach}

We extracted keyphrases, which are automatically extracted keywords, as per the methodology described in Ferrera’s work\cite {Ferrara2011} from the title of all documents, as well as from the title and abstract of the 3.3 million documents with English abstracts. The extraction process of keyphrases is a multi-step process using the open-source Distiller framework[22]. Distiller provides an easy-to-use pipeline to automate the process of extracting keyphrases by specifying the steps in the pipeline. A flowchart of the steps can be seen in  Fig. 1. The title and the abstract, if used, of each document is tokenized, POS-tagged, stripped of stop words, and stemmed using the Porter Stemmer to generate a stream of tokens. From these tokens, continuous combinations which match pre-specified POS patterns are selected as candidate keyphrases. 
\begin{figure}
    \centering
    \includegraphics[scale=1.5]{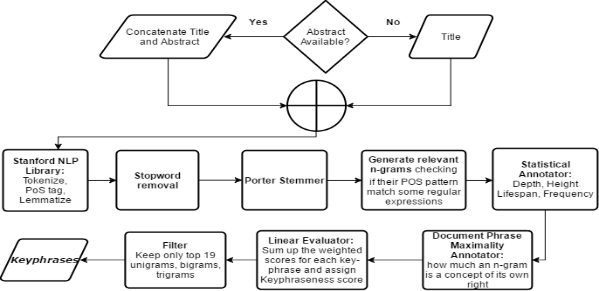}
    \caption{Steps involved in constructing keyphrases from a document}
    \label{fig:4.5}
\end{figure}

For instance, two pre-specified patterns are “NN” and "NN/NN/NN”, which stands for ‘Noun, singular or mass’ and ‘a sequence of three nouns’ respectively. Consider an example title: “Research Paper Recommender Systems – A quantitative study of performance”. Table 4.5 shows the results of processing this title through the pipeline and Table 4.6 presents a few candidate keyphrases that can be extracted from this title.

\begin{figure}
    \centering
    \includegraphics[scale=0.5]{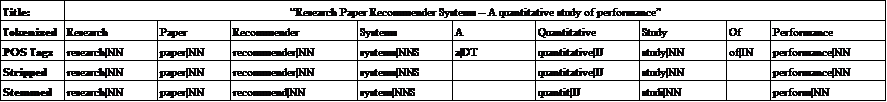}
    \caption{Example of Steps in Keyphrase Extraction Pipeline}
    \label{fig:4.6}
\end{figure}

Next, each candidate is annotated with statistical information such as the depth in the document at which it was extracted from, the portion of the document which is encompassed by the first and last occurrence of the keyphrase, called the lifespan, and the number of occurrences of the keyphrase. Each candidate is then given a score corresponding to a concept known as document phrase maximality, which expresses how much that candidate keyphrase is a concept of its own right.

\begin{figure}
    \centering
    \includegraphics[scale=0.5]{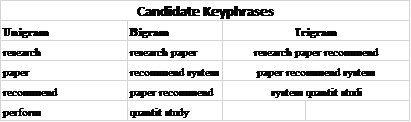}
    \caption{Examples of Candidate keyphrases }
    \label{fig:4.7}
\end{figure}

\begin{figure}
    \centering
    \includegraphics[scale=0.5]{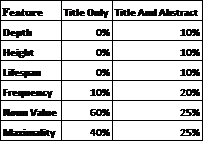}
    \caption{Weights used to score candidate keyphrases}
    \label{fig:4.8}
\end{figure}
Finally, these scores are weighted and summed up to constitute a final score known as a keyphraseness score as shown in Fig. 4.7. We used different weights, chosen apriori, to extract keyphrases from titles only as opposed to keyphrases from the title and abstract. The original work[6] used an apriori cutoff for the keyphrase-ness score to choose keyphrases. In contrast, we filtered out 19 keyphrases in each of the unigram, bigram, and trigram categories in descending order of keyphraseness score. These extracted keyphrases are used to calculate the similarity between docu-ments. To calculate document similarities, we indexed the each document’s keyphrases using Lucene 6.3.0. 

\chapter{Results} 

\label{Chapter5} 

\lhead{Results} 


\section{Data Collection}

We recorded 3.6 million requests for recommendations, approximately 35,000 requests for recommendations yielding 210,000 recommendations per day. In total, we received 7700 clicks on these recommended documents. The CTR was 0.21\%.  

\section{Comparison of Baselines:}

We first compared the performance of the suggested baselines without accounting for the difference in language of the input document. 
As illustrated in Fig.1, in the first case, Lucene’s MLT approach, with a CTR of 0.229\% outperforms every other approach. It was surprising that the Keyphrase approach (0.148\%) did not  do better than even the random approach, with and without the language restriction (0.149\%, 0.159\%). The stereotype approach had a strong performance with CTR of 0.194\%, confirming the usefulness of incorporating stereotypes as a baseline.

Since the initial results did not bring clarity to the usefulness of the Keyphrase approach, we repeated the analysis considering only recommendations for which the input document had an English title. As seen in Fig.1, in the second case, it cannot be said with statistical significance that Lucene’s MLT with a CTR of 0.169\% outperforms the key-phrase approach (CTR=0.148\%). The recommendation performance of the suggested approaches also depends on the type of user requesting recommendations. This inference could explain the observation of Beel et al. that the performance of an approach varies from corpus to corpus\cite{beel}. For example, an explanation for these results might be that German users are more likely to click on a recommendation than users looking exclusively at English documents. It should also be noted that the Stereotype approach maintained its CTR of 0.204\%. 

On further study of the stereotype and most-popular recommendations over a longer period of time and 28 million delivered recommendations, it was found that Most-popular recommendations achieved a CTR of 0.11\%, and stereotype recommendations achieved a CTR of 0.124\%. Compared to a “random recommendations” baseline (CTR 0.12\%), and a content-based filtering baseline (CTR 0.145\%)\cite{beel2017stereotype}.

\section{Comparison By Language}
We further drilled down into the data by comparing only between recommendation requests for documents with English titles and English abstracts. In this case, the results favored the keyphrase approach (0.170\%) over the other CBF approach Lucene MLT (0.120\%), as well as the random approaches. We infer that the keyphrase approach does a better job at removing the clutter from the abstracts while retaining the meaning of the text when compared to Lucene’s MLT. Again, the stereotype approach performed well in relative terms with a CTR of 0.178\%. 

We cannot yet draw definitive conclusions to separate the two random approaches because we do not have enough data yet. However, preliminary data shows that it can serve as a useful approach to evaluate any other approach against because of its steady performance and ease of implementation. Finally, the differences in the CTRs just because of the difference in the language of the input documents supports our reasoning behind using a set of baselines as opposed to just one or two.

\begin{figure}
    \centering
    \includegraphics[scale=0.85]{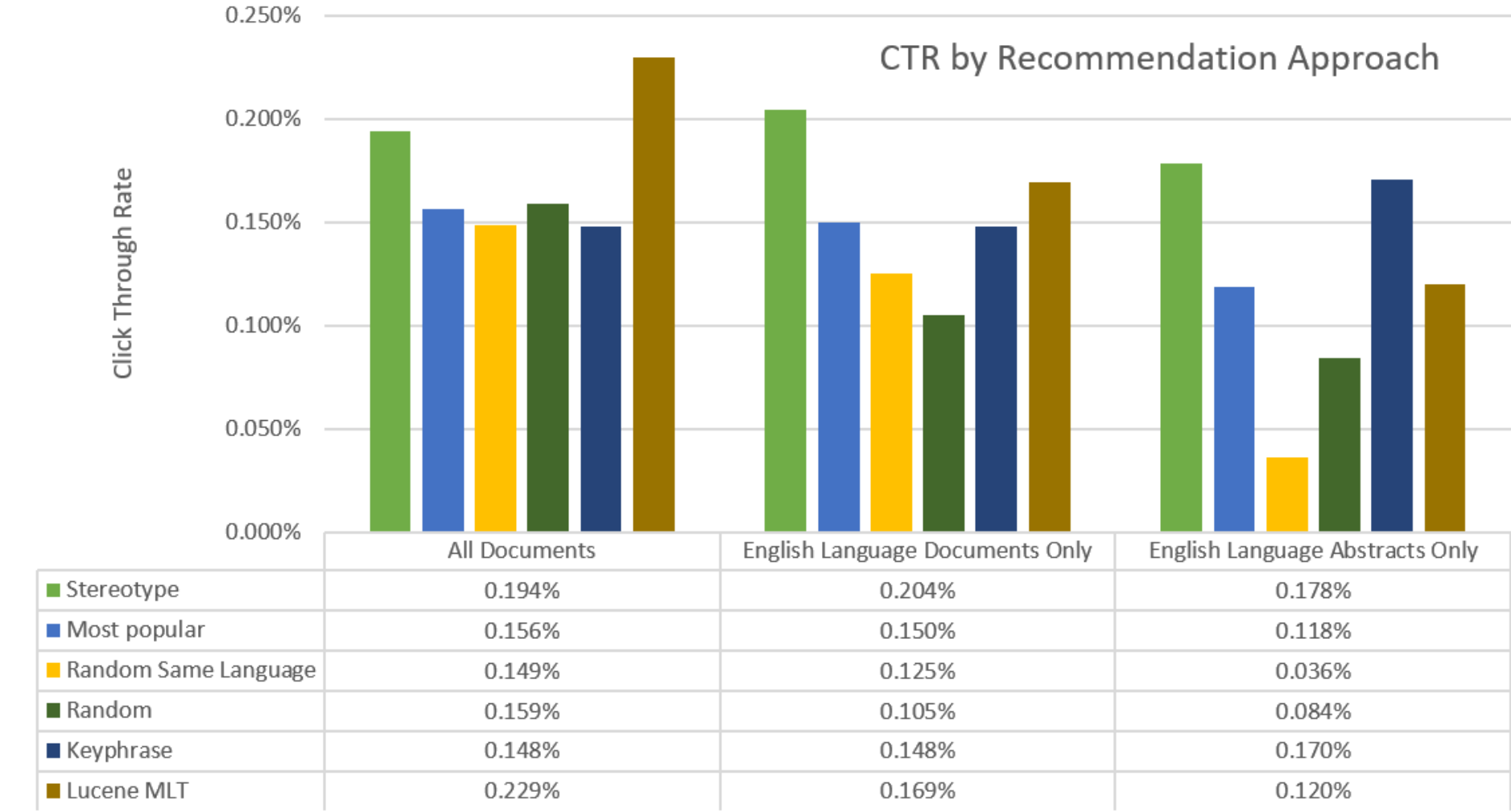}
    \caption{Comparisons of Approaches by Recommendation Request Type}
\end{figure}

\section{Analysis of Keyphrase Approach}

\subsection{Overall Comparison}
The CTR for all 31 million recommendations Mr. DLib delivered in the period be-tween 17th October, 2016 and 17th January, 2017 is 0.138\%. Of these, close to 24 million recommendations were made using Lucene’s MLT function at a CTR of 0.147\%. These were primarily recommendations for German language documents. 

The right two columns in Fig. 5.2 illustrate the number of delivered recommendations and CTRs for recommendations delivered by Mr. DLib related to English documents. The keyphrase algorithm (CTR = 0.067\%) performed worse than the Lucene MLT implementation (CTR = 0.085\%). However, the keyphrase algorithm did perform better than the completely random recommendations (0.055\%).

\begin{figure}
    \centering
    \includegraphics[scale=0.7]{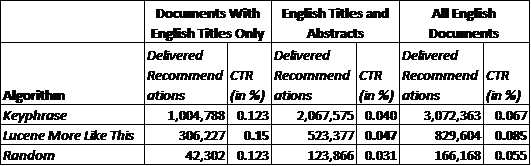}
    \caption{Overall Comparison of Recommendation Algorithms}
\end{figure}

Fig. 5.2 also presents the split-up between the recommendations that Mr. DLib delivered for documents which had English titles only, as opposed to documents which had English titles and abstracts. There is a big difference in CTRs between corre-sponding algorithms in the two categories, for instance, the Keyphrase algorithm rec-orded a CTR of 0.123\% with recommendations for documents with just the title in English, whereas the CTR (=0.040\%) dropped for documents having English titles and abstracts. In both cases, the Lucene MLT implementation (CTR=0.15\% and 0.047\%) outperformed the keyphrase algorithm (CTR=0.123\% and 0.040\%). While the outlook seems better when the keyphrase algorithm is used only on English titled documents, it did not outperform the random baseline which also had a CTR of 0.123\%.

\subsection{Inclusion of Abstract}
Fig. 5.3 presents the number of delivered recommendations, click counts and CTRs for Keyphrase based recommendations made for documents which had English ab-stracts available. There is no statistically significant difference in the CTRs, 0.409\% when we used only the title, and 0.393\\

\begin{figure}
    \centering
    \includegraphics[]{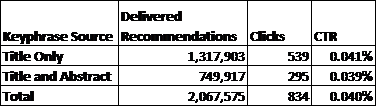}
    \caption{Comparison of Keyphrase Algorithm based on Source of Keyphrases}
\end{figure}

\subsection{Effect of N-gram Type}
As there were noticeable differences between the CTRs when documents had an English abstract available compared to documents with only an English title, we split the analysis by the type of extracted keyphrase into two sections. The left side of  Fig 5.4 compares the CTR of recommendations provided using different combina-tions of n-grams. The results show that no one particular combination of n-gram was more effective than the others. The range of CTRs was 0.012, with the lowest being Trigrams at 0.068\% and the highest being a combination of Unigrams and Bigrams at 0.0803\%. When we provided recommendations for documents having an abstract, the average CTRs were lower, as described in Section 5.4.2. The results again do not indicate that one combination of n-grams had any advantage over the others. 

\begin{figure}
    \centering
    \includegraphics[]{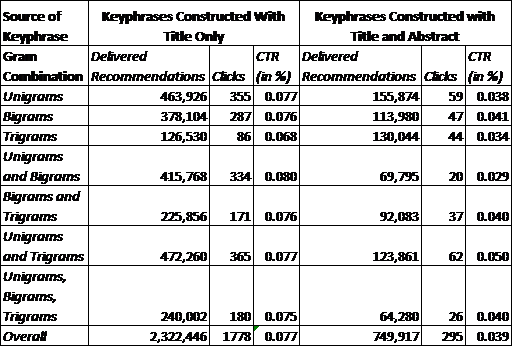}
    \caption{CTR comparison based on the n-gram type of extracted keyphrase}
\end{figure}

\subsection{Effect of Keyphrase Count}

Table 6 outlays the experimental results for recommendations that used differing number of keyphrases to compare the similarity between documents which shall be termed keyphrase\_count henceforth. We grouped the counts for comparative pur-poses as the data was sparse and skewed toward lower keyphrase\_counts. Once again, we have split the comparison based on whether the query document had an abstract or not. 

On the left hand side of the table, we display the results from the recommendations provided using keyphrases built out of just the title. As was expected, with most titles being between 5 and 10 words, there were not too many recommendations delivered with keyphrase\_count > 10. Most recommendations were made with a keyphrase\_count of 1 or 2 only. Additionally, the sparsity in the data when keyphrase\_count > 10 meant that the high CTRs recorded in these cases might be anomalous. There is also no significant difference between the CTRs up to keyphrase\_count = 5.

\begin{figure}
    \centering
    \includegraphics[]{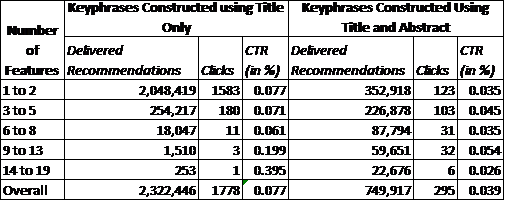}
    \caption{Comparison of CTRs based on number of keyphrases used for document similarity}
\end{figure}

In the second case, where documents had abstracts available, the distribution of keyphrase\_count is more uniform because more number of candidate keyphrases are available. We observe that although the CTR is highest when keyphrase\_count is between 9 and 13 (0.054\%), the difference is not statistically significant as compared to the CTR when 3 < keyphrase\_count < 5 (0.045\%). However, both these settings perform better than when keyphrase\_count < 3 (0.035\%), or keyphrase\_count be-tween 5 and 8 (0.035\%). 

\chapter{Future Work} 

\label{Chapter6} 

\lhead{Future Work} 


Through the study of the data we collected in this research, we understood that there are significant variations in the click through rate of recommendations based on the language of the document for which we provided the recommendations. This means that there is a possible scope to improve click through rates by providing cross language recommendations of literature documents. 

First, we will be focusing on the German-English language barrier. The steps in this process include:
\begin{enumerate}
    \item Identify tools and frameworks to translate abstracts, titles, keywords to English
    \item Translate documents and store in our database
    \item Calculate keyphrases for translated documents
    \item Re-index Solr (both for keyphrases and translated titles, abstracts, keyphrases for Lucene MLT recommender)
    \item Rewrite code to experiment with cross-language recommendations.
\end{enumerate}

We have already identified that there are two or three types of machine translation frameworks. These include Rule Based Machine Translation, Statistical Machine Translation and a hybrid of these techniques. Usually, rule based machine translation does not require much processing and the translations thus formed would generally lack in grammatical correctness. 

Statistical machine translation, on the other hand, has generally been proven to be more effective in the accuracy of the translation in the sense of the grammar and the meaning of the sentence. Thus, for the purpose of literature recommendations, it would be interesting to study which of these two methods provide the better result in terms of click through rates. Of course, CTR might not be the ideal metric that should be used for comparison. Further study has to be conducted in this area to identify relevant metrics for this field.

\chapter{Summary} 

\label{Chapter8} 

\lhead{Summary} 


The ever expanding digitization of research literature has created a need for efficient search, information retrieval and recommendation approaches in order to correctly identify and distribute relevant and personalized research documents to researchers. After identifying the need to progress the state of the art in the area of research paper recommendation for digital libraries, this thesis described the existing literature in this respect over the last four years. This was an addition to Beel's extensive literature survey for this field \cite{Beel:2013:CAO:2532508.2532511}.

The literature survey in Chapter \ref{Chapter 2} identified the lack of a common set of baselines to compare incremental innovations in related article search for research documents. This was a stumbling block in the speed with which the state of the art advanced in this area. In Chapter \ref{Chapter3}, we set out the methodology for this thesis. The basis of the experiments was described by identifying the features that we thought to be relevant in classifying a baseline as a "good" one or not. With these features in mind, five different recommendation approaches were selected and described that satisfied the aforementioned criterion. Finally, we explained the procedure of the experiment and the idea behind it.

In Chapter \ref{Chapter4}, the implementation of the platform for the experiment, Mr. DLib, was illustrated and explained. Knowing the architecture of the system is a good way to source feedback and criticism to improve the performance of the system. Chapter \ref{Chapter4} also addressed the process of a literature recommendation and showcased the format by which these literature recommendations were displayed on our partner's digital library. 

Chapter \ref{Chapter5} talks about the results of the experiment and the conclusions that we were able to draw from it. Although not entirely conclusive, the results were a great starting point in illustrating the need for a common set of baselines. It was observed that the Click Through Rates (CTR) differed not only by the choice of recommendation approach, but also by the inherent characteristics of the document for which we were recommending literature. 

One such characteristic that affected the CTR was the language of the original document. We observed that Sowieport being a German based digital library, there was an overall higher CTR for German documents, and the converse was true for English documents. Thus, we believe that to fully understand the performance of an approach, particularly in a multilingual digital library such as Sowieport, it is important to make our recommendations cross language boundaries. 

Thus, in Chapter \ref{Chapter6} we set out the future work which involves implementing Cross Language literature recommendations on Mr. DLib. This involves the work of translating abstracts and titles in order to make it possible to identify similar documents across language boundaries. There is also a focus on implementing citation based recommendations, provided there is access to document citations and references. As this is an experiment in a live setting, the results would become more and more concrete as time passes.



\addtocontents{toc}{\vspace{2em}} 




\addtocontents{toc}{\vspace{2em}} 

\backmatter


\label{Bibliography}

\lhead{\emph{References}} 

\bibliographystyle{ieeetr}

\bibliography{Bibliography}

\end{document}